\def\ra{\rangle}

\def\be{\begin{equation}}
\def\ee{\end{equation}}
\def\ba{\begin{array}}
\def\ea{\end{array}}

\def\qed{\leavevmode\unskip\penalty9999 \hbox{}\nobreak\hfill
     \quad\hbox{\leavevmode  \hbox to.77778em{%
               \hfil\vrule   \vbox to.675em%
               {\hrule width.6em\vfil\hrule}\vrule\hfil}}
     \par\vskip3pt}

\documentclass[amsmath,preprint,amssymb,amsfonts,smallextended]{revtex4}
\usepackage{epsfig}

\newtheorem{theorem}{Theorem}

\begin{document}
\begin{center}
\bf{Entanglement Detection and Distillation for Arbitrary Bipartite Systems}
\end{center}

\begin{center}
Ming-Jing Zhao$^{1}$, Ting-Gui Zhang$^{1}$, Xianqing Li-Jost$^{1}$ and Shao-Ming Fei$^{1,2}$

\vspace{2ex}

{\small $~^{1}$ Max-Planck-Institute for Mathematics in the Sciences,
Leipzig 04103, Germany}

{\small $~^{2}$ School of Mathematical Sciences, Capital Normal
University, Beijing 100048, China}

\end{center}


{{\bf Abstract}
We present an inequality for detecting entanglement and distillability
of arbitrary dimensional bipartite systems. This inequality provides
a sufficient condition of entanglement for bipartite mixed states,
and a necessary and sufficient condition of entanglement for
bipartite pure states. Moreover, the inequality also gives a
necessary and sufficient condition for distillability.}


\section{Introduction}

Entanglement is one of the most fascinating features of quantum
theory and has numerous applications in quantum information
processing \cite{M.A.Nielsen}. As a result, various approaches have
been proposed and many significant conclusions have been derived in
detecting entanglement \cite{O. Guhne,Peres A., M. Horodecki, M.
Lewenstein, M. Horodecki1999, M. Nielsen,P. Horodecki1997,K.Chen,O.Rudolph,D.
Cavalcanti,p1,p2}. However there are yet no operational necessary and sufficient separability criteria for general higher dimensional
quantum states. In particular, experimental detection of quantum
entanglement by measuring some suitable quantum mechanical
observables has practical importance. The Bell inequalities
can be used to detect perfectly the entanglement of pure
bipartite states \cite{CHSH,N. Gisin92,D. Collins, J. Chen, M. Li}.
Besides Bell inequalities, the entanglement witness could also be
used for experimental detection of quantum entanglement for some
special states \cite{M. Horodecki, Philipp05, M. Lewenstein01},
which could give rise to entanglement estimation \cite{J. Eisert}.
The two-copy measurement of concurrence for two-qubit pure states
has been already realized experimentally \cite{F. Mintert2005,S. P.
Walborn2006,S. P. Walborn2007}. For arbitrary dimensional bipartite
pure states, one method of measuring concurrence has been presented
in terms of one-copy measurement \cite{mj1}. Except that, uncertainty relations are also favorable in entanglement detection, for instance, one separability condition has been derived for all negative partial transpose state in experimentally accessible forms \cite{H. Nha}.

For bipartite mixed states, a necessary and sufficient inequality
has been derived for detecting entanglement of two-qubit states
\cite{sixiayu}. In Ref. \cite{mj} based on a different approach, an
inequality is presented, which is both necessary and sufficient in
detecting entanglement of qubit-qutrit states, and necessary for
qubit-qudit states.

In this article we propose an inequality detecting entanglement for
arbitrary dimensional bipartite states, which consists of local
observables with measurement outcomes $\pm 1$. This inequality gives
a necessary condition of separability for mixed states, namely, any
violation of the inequality implies entanglement. It is shown that
our inequality can detect entanglement of a wide class of states such
as Horodecki's state, isotropic state and Werner state. For pure
states, the inequality becomes both necessary and sufficient. All
pure entangled states violate it.
Moreover, our inequality provides also a necessary and sufficient condition for entanglement distillability, an approach to
get the ideal resource from general quantum mixed states for quantum
information processing \cite{C. H. Bennett1996,M. Horodecki1998}.

The paper is organized as follows. In Sec. II, we first propose an
inequality in terms of the mean values of local observables. Then we
prove that all separable states obey this inequality and all pure
entangled states violate it. As applications some detailed examples
are presented. We show then that the inequality is a necessary and sufficient condition for distillability. Comments and conclusions are
given in the last section.

\section{Inequality for arbitrary bipartite systems}

Let $H_m$, $H_n$ be $m$, $n$-dimensional vector spaces with
$\{|i\rangle\}_{i=0}^{m-1}$ and $\{|j\rangle\}_{j=0}^{n-1}$ the
computational basis respectively. Set $\lambda_0^A=I_m$ the $m\times m$ identity matrix,
$\lambda_i^A=|0\rangle\langle0| - |i\rangle\langle i|$,
$\mu_1^A=|0\rangle\langle1| + |1\rangle\langle0|$, and
$\mu_2^A= {\rm i}|0\rangle\langle1| -{\rm i} |1\rangle\langle0|$, where $|i\ra\in
H_m$, $i=1,\cdots,m-1$. Set $\lambda_0^B=I_n$ the $n\times n$ identity matrix, $\lambda_i^B=|0\rangle\langle0| -
|i\rangle\langle i|$, $\mu_1^B=|0\rangle\langle1| +
|1\rangle\langle0|$, and $\mu_2^B={\rm i}|0\rangle\langle1| -{\rm i}
|1\rangle\langle0|$, where $|i\ra\in H_n$, $i=1,\cdots,n-1$. Let
$A_i= U\lambda_i^A U^\dagger$, $i=0,1,\cdots, m-1$, $A_j^\prime=U
\mu_j^A U^\dagger$, $j=1,2$, be a set of quantum mechanical
observables acting on the first subsystem, with $U$ any $m\times m$
unitary matrix. Let $B_i= V\lambda_i^B V^\dagger$, $i=0,1,\cdots,
n-1$, $B_j^\prime=V \mu_j^B V^\dagger$, $j=1,2$, be a set of quantum
mechanical observables acting on the second subsystem, with $V$ any
$n\times n$ unitary matrix. Here Roman letter $\rm i$ represents the imaginary unit.

We define
\begin{widetext}
\be\label{operators}
\ba{rcl}
\displaystyle H^{(m,n)}_{U,V}=&&\displaystyle \frac{1}{mn} \sum_{i, j}(1 - \frac{m}{2}\delta_{i1}-\frac{n}{2}\delta_{j1}) A_i\otimes B_j ;\\
\displaystyle P^{(m,n)}_{U,V}=&&\displaystyle \frac{1}{2m^2n^2}
\sum_{i,j}(m\delta_{i1}-n\delta_{j1}) A_i\otimes B_j ;\\
\displaystyle Q^{(m,n)}_{U,V}=&&\displaystyle \frac{1}{16}( A_1^\prime \otimes B_1^\prime -A_2^\prime \otimes
B_2^\prime),
\ea
\ee
\end{widetext}
where $\delta_{kl}=1$ if $k=l$ and is zero otherwise. According to
the mean values of the operators in Eq. (\ref{operators}), we can
construct an inequality detecting entanglement for $m\otimes n$
systems.

\begin{theorem}
Any separable state $\rho$ in $H_m\otimes H_n$ obeys the following inequality
\begin{eqnarray}\label{ineq nbyn}
\langle H^{(m,n)}_{U,V}\rangle_{\rho}^2 \geq\langle
P^{(m,n)}_{U,V}\rangle^2_{\rho}+ \langle Q^{(m,n)}_{U,V}\rangle^2_{\rho}
\end{eqnarray}
for all $m\times m$ unitary matrix $U$ and $n\times n$ unitary matrix $V$.
\end{theorem}

Proof. First we show that inequality
(\ref{ineq nbyn}) holds for all pure separable states, which is
equivalent to prove that for arbitrary pure separable state $\rho$,
the following inequality holds:
\begin{eqnarray}\label{ineq nbyn*}
\langle H^{(m,n)}_{I_m,I_n}\rangle_{\rho}^2 \geq\langle
P^{(m,n)}_{I_m,I_n}\rangle^2_{\rho}+ \langle Q^{(m,n)}_{I_m,I_n}\rangle^2_{\rho}.
\end{eqnarray}

Note that any pure separable state can be written as $|\xi\rangle= \sum_{i=0}^{m-1}\sum_{j=0}^{n-1}a_ib_j|ij\rangle$ with $\sum_{i=0}^{m-1} |a_i|^2=\sum_{j=0}^{n-1}|b_j|^2=1$.
Inserting this separable pure state $\rho=|\xi\rangle \langle \xi|$ into Eq. (\ref{ineq nbyn*}),
one gets that
\begin{eqnarray}\label{proof pure sep left}
\langle H^{(m,n)}_{I_m,I_n}\rangle_{|\xi\rangle \langle \xi|}
=\frac{1}{2}(|a_0b_1|^2+|a_1b_0|^2)\geq0.
\end{eqnarray}
While the right hand
side of inequality (\ref{ineq nbyn*}) becomes
\begin{eqnarray}\label{proof pure sep right}
\langle P^{(m,n)}_{I_m,I_n}\rangle^2_{|\xi\rangle \langle \xi|}+ \langle Q^{(m,n)}_{I_m,I_n}\rangle^2_{|\xi\rangle \langle \xi|}
=\frac{1}{4}(|a_0b_1|^2-|a_1b_0|^2)^2+Re(a_0a_1^*b_0b_1^*)^2.
\end{eqnarray}
From Eqs. (\ref{proof pure sep left}) and (\ref{proof pure sep right}), it is easy to obtain that inequality (\ref{ineq nbyn*}) holds for any pure separable states.

We now prove that inequality (\ref{ineq nbyn})
also holds for general separable mixed states,
$$\rho=\sum_i p_i |\psi_i\rangle \langle \psi_i|,~~~0\leq p_i\leq 1,~~~\sum_i p_i =1,$$
where
$|\psi_i\rangle$ are all pure separable states. Denote
\begin{eqnarray*}
c_i&&=\langle H^{(m,n)}_{U,V}\rangle_{|\psi_i\rangle\langle \psi_i|};\\\nonumber
d_i&&=\langle P^{(m,n)}_{U,V}\rangle_{|\psi_i\rangle\langle \psi_i|};\\\nonumber
e_i&&= \langle Q^{(m,n)}_{U,V}\rangle_{|\psi_i\rangle\langle \psi_i|}.
\end{eqnarray*}
Since inequality (\ref{ineq nbyn}) holds for all pure separable states,
one has $c_i^2 \geq d_i^2 +e_i^2$.
Taking into account that if the inequality
$c_i^2 \geq d_i^2 +e_i^2$ holds for arbitrary real numbers $d_i$,  $e_i$ and nonnegative
$c_i$, $i=1, \cdots, N$, then
$(\sum_{i=1}^N p_i c_i)^2 \geq (\sum_{i=1}^N p_i d_i)^2
+(\sum_{i=1}^N p_i e_i)^2$
for $0 \leq p_i \leq 1$ and $\sum_{i=1}^N p_i=1$,
we know that any mixed
separable state $\rho$ obeys inequality (\ref{ineq nbyn}).\qed

This theorem shows that any violation of this inequality implies
entanglement. Hence it gives a sufficient condition of entanglement
for arbitrary dimensional bipartite systems. Next we prove that for
the pure state case, one has that all pure entangled states violate
this inequality for some unitary $U$ and $V$, so that the inequality
provides a necessary and sufficient condition for entanglement of
pure states.

\begin{theorem}\label{th pure violate}
A pure state in $H_m\otimes H_n$ is entangled if and only if it
violates inequality (\ref{ineq nbyn}) for some unitary matrices
$U$ and $V$.
\end{theorem}

Proof. Here we only need to prove that any pure entangled state in
$H_m\otimes H_n$ violates inequality  (\ref{ineq nbyn}) for some
unitary matrices $U$ and $V$. Without loss of generality, we assume
$m\geq n$. By Schmidt decomposition any entangled pure state
$|\phi\rangle$ can be transformed into $|\phi^\prime\rangle$:
\be\label{t2p} |\phi^\prime\rangle=U^\dagger\otimes V^\dagger
|\phi\rangle=\sum_{i=0}^{n-1} a_i|ii\rangle, \ee with $a_0>0$,
$a_1>0$, and $a_i\geq0$ for $i=2,\cdots,n-1$, $\sum_{i=0}^{n-1}
|a_i|^2=1$. Here $U$ is an $m\times m$ unitary matrix and $V$ is an
$n\times n$ unitary matrix. Then from the PPT criterion \cite{Peres
A., M. Horodecki} there exist nonzero real numbers $a$ and $b$
satisfying $a^2+b^2=1$ and $|\psi\rangle=a|01\rangle +b|10\rangle$
such that
\begin{eqnarray*}
\langle (|\psi\rangle \langle\psi|)^{T_1} \rangle_{|\phi^\prime\rangle\langle\phi^\prime|}
=\langle \psi| (|\phi^\prime\rangle \langle\phi^\prime|)^{T_1} |\psi\rangle
<0.
\end{eqnarray*}
By expanding the partial transposed matrix $|\psi\rangle \langle
\psi|^{T_1}$ according to the matrices $\{\lambda_i^A\}$,
$\{\mu_i^A\}$, $\{\lambda_i^B\}$ and $\{\mu_i^B\}$ on the first and
second subsystems respectively, we get
\begin{widetext}
\begin{eqnarray}\label{nbyn pure pt}
&&\langle (|\psi\rangle \langle\psi|)^{T_1} \rangle_{|\phi^\prime\rangle\langle\phi^\prime|}\\\nonumber
&&=\frac{1}{mn}\langle  \sum_{i, j\neq 1} \lambda_i\otimes \lambda_j + \sum_{i,j}[ (\frac{-m+2}{2}+\frac{m}{2}\sqrt{1-C^2})\delta_{i1}\\\nonumber
&&+(\frac{-n+2}{2}-\frac{n}{2}\sqrt{1-C^2})\delta_{j1}] \lambda_i\otimes \lambda_j  \rangle_{|\phi^\prime\rangle\langle\phi^\prime|}  + \frac{1}{4}C\langle \mu_1^A \otimes \mu_1^B -\mu_2^A \otimes \mu_2^B\rangle_{|\phi^\prime\rangle\langle\phi^\prime|}\\\nonumber
&&\geq \langle H^{(m,n)}_{I_m,I_n}\rangle_{|\phi^\prime\rangle\langle\phi^\prime|}
-\left|\sqrt{1-C^2}\langle P^{(m,n)}_{I_m,I_n}\rangle_{|\phi^\prime\rangle\langle\phi^\prime|}
+ C\langle Q^{(m,n)}_{I_m,I_n}\rangle_{|\phi^\prime\rangle\langle\phi^\prime|}\right| \\\nonumber
&&\geq\langle H^{(m,n)}_{I_m,I_n}\rangle_{|\phi^\prime\rangle\langle\phi^\prime|}
-\left(\langle P^{(m,n)}_{I_m,I_n}\rangle^2_{|\phi^\prime\rangle\langle\phi^\prime|} + \langle Q^{(m,n)}_{I_m,I_n}\rangle^2_{|\phi^\prime\rangle\langle\phi^\prime|}\right)^{\frac{1}{2}}\\\nonumber
&&=\langle H^{(m,n)}_{U,V}\rangle_{|\phi\rangle\langle\phi|}
-\left(\langle P^{(m,n)}_{U,V}\rangle^2_{|\phi\rangle\langle\phi|} + \langle Q^{(m,n)}_{U,V}\rangle^2_{|\phi\rangle\langle\phi|}\right)^{\frac{1}{2}},
\end{eqnarray}
\end{widetext}
where $C=2ab$ is just the concurrence of the pure state
$|\psi\rangle$, defined by $C(|\psi \rangle)= \sqrt{2(1-Tr
\rho_1^2)}$. $\rho_1$ is the reduced density matrix
$\rho_1=Tr_2(|\psi \rangle \langle \psi|)$, where $Tr_2$ stands for
the partial trace with respect to the second subsystem. Here $A_i= U
\lambda_i^A U^\dagger$, $i=0,1,\cdots, m-1$, $A_j^\prime=U \mu_j^A
U^\dagger$, $j=1,2$, $B_i= V \lambda_i^B V^\dagger$, $i=0,1,\cdots,
n-1$, $B_j^\prime=V \mu_j^B V^\dagger$, $j=1,2$. $U$ and $V$ are
just the unitary matrices from the Schmidt decomposition in Eq.
(\ref{t2p}). The first inequality is due to $-|x|\leq x$ and the
second one is from the Cauchy inequality. Since the left hand side
of (\ref{nbyn pure pt}) is negative, we have that  the pure
entangled state $|\phi\rangle$ violates inequality (\ref{ineq
nbyn}) with respect to the corresponding observables. \qed

Therefore any pure state $|\psi\rangle$ in $H_m\otimes H_n$ is separable if and only
if inequality (\ref{ineq nbyn}) is satisfied.
Inequality (\ref{ineq nbyn}) gives a necessary and sufficient criterion
for the separability of pure states, which may be determined by experimental measurements
on the local observables. For example, for pure entangled state $|\phi^\prime\rangle$ defined above,
it violates inequality (\ref{ineq nbyn*}) corresponding to $U=I_m$ and $V=I_n$ in inequality (\ref{ineq nbyn}).

In fact for pure bipartite states there are already many Bell inequalities like
Refs. \cite{N. Gisin92,M. Li} in terms of the expectation of the local observables and
Refs. \cite{D. Collins,J. Chen} in terms of probabilities for $d\otimes d$ systems.
Inequality (\ref{ineq nbyn}) is for arbitrary dimensional pure bipartite systems, no
matter whether the dimensions of the two subsystems are the same or not. Moreover,
it also provides sufficient condition for entanglement of mixed states.

Inequality (\ref{ineq nbyn}) can be associated with nonlinear
entanglement witness operators, where the violation is replaced by a
negative expectation value. In fact, for given $U$ and $V$,
inequality (\ref{ineq nbyn}) gives rise to a kind of nonlinear
entanglement witness $W^{(m,n)}_{U,V}$:
\begin{eqnarray*}
\langle W^{(m,n)}_{U,V}\rangle_{\rho}\equiv\langle H^{(m,n)}_{U,V}\rangle^2_{\rho}-(\langle
P^{(m,n)}_{U,V}\rangle^2_{\rho}+ \langle Q^{(m,n)}_{U,V}\rangle^2_{\rho}).
\end{eqnarray*}
For all separable states $\sigma$, $\langle
W^{(m,n)}_{U,V}\rangle_{\sigma}\geq 0$. If $\langle
W^{(m,n)}_{U,V}\rangle_{\rho}<0$ then $\rho$ is entangled. As an
entanglement witness gives rise to a superface in the state space,
dividing the space into a set of entangled states and the rest, the
detection of different entangled states depends on the choice of
different witnesses. Here different choices of $U$ and $V$ detect
different sets of entangled states.

Generally linear entanglement witness
operators can be associated with positive maps in terms of
Jamiolkowski isomorphism. And positive maps could give rise to
entanglement linear witness operators, see e.g. \cite{qh}.
Our inequality (\ref{ineq nbyn}) is not linear, as many
Bell-type inequalities for mixed states \cite{sixiayu,mj}.
The entanglement witness operators deduced from the
inequality is also nonlinear. The direct relations between the
entanglement witness operators and the positive maps are not obvious.

Subsequently, we consider the maximal violation of inequality
(\ref{ineq nbyn}). Let $F^{(m,n)}(\rho)=\max_{U,V}\{-\langle
W^{(m,n)}_{U,V}\rangle_{\rho},0\}$ denote the maximal violation
value with respect to a given state $\rho$, under all $U$ and $V$.
Then $F^{(m,n)}(\rho)=0$ if $\rho$ is separable. Additionally,
$F^{(m,n)}(\rho)$ is invariant under the local unitary
transformations.

From another point of view, inequality (\ref{ineq nbyn}) can be
viewed as the generalization of the main result in Ref.
\cite{sixiayu} and Ref. \cite{mj}. For $m=n=2$, inequality
(\ref{ineq nbyn}) becomes the one in Ref. \cite{sixiayu} which is
necessary and sufficient in detecting entanglement for pure or mixed
states. For $m=2$ and $n=3$, inequality (\ref{ineq nbyn}) is
also the necessary and sufficient condition of separability
\cite{mj}. But for $m>3$ or $n>3$, this inequality is only necessary
for separability of mixed states, but necessary and sufficient for
pure states. As is shown in Ref. \cite{mj}, for $m=2$ the inequality
can detect some PPT entanglement.
In the following we give some examples concerning entanglement
detection in terms of inequality (\ref{ineq nbyn}).

Example 1. Horodecki's $3\otimes 3$ state:
\begin{eqnarray*}
\sigma_\alpha = \frac{2}{7}|\psi^+\rangle \langle \psi^+| +\frac{\alpha}{7}\sigma_+ +\frac{5-\alpha}{7}\sigma_-,
\end{eqnarray*}
where $\sigma_+=\frac{1}{3}(|01\rangle \langle 01| + |12\rangle
\langle 12| + |20\rangle \langle 20|)$,
$\sigma_-=\frac{1}{3}(|10\rangle \langle 10| + |21\rangle \langle
21| + |02\rangle \langle 02|)$,
$|\psi^+\rangle=\frac{1}{\sqrt{3}}(|00\rangle + |11\rangle +
|22\rangle)$. $\sigma_\alpha$ is separable for $2\leq \alpha \leq
3$, bound (PPT) entangled for $3< \alpha \leq 4$, and  free
entangled for $4< \alpha \leq 5$ \cite{P. Horodecki}. Employing inequality (\ref{ineq nbyn*}) for $3\otimes 3$ systems,
\begin{equation}\label{ineq 3by3*}
\begin{array}{rcl}
&&4\langle 2I_3 \otimes I_3 - I_3 \otimes \lambda_1^B + 2I_3 \otimes \lambda_2^B
-\lambda_1\otimes I_3-4\lambda_1\otimes \lambda_1^B \\[2mm]
&&- \lambda_1^A\otimes \lambda_2^B
+2\lambda_2^A\otimes I_n-\lambda_2^A\otimes \lambda_1^B + 2\lambda_2^A\otimes \lambda_2^B\rangle^2_{\rho} \\[2mm]
&&\geq36\langle
I_3 \otimes \lambda_1^B -\lambda_1^A \otimes I_3 -\lambda_1^A \otimes \lambda_2^B +\lambda_2^A\otimes \lambda_1^B \rangle^2_{\rho}
\\[2mm]
&& + 81\langle \mu_1^A \otimes \mu_1^B +\mu_2^A \otimes
\mu_2^B\rangle^2_{\rho},
\end{array}
\end{equation}
we have that the left hand side of inequality (\ref{ineq 3by3*})
is $\frac{900}{49}$ and the right hand side of inequality
(\ref{ineq 3by3*}) is $\frac{36(2\alpha-5)^2}{49}+\frac{576}{49}$.
Hence $\sigma_{\alpha}$ violates inequality (\ref{ineq 3by3*})
if and only if $\alpha>4$.

Example 2.
Isotropic states are a class of $U \otimes
U^*$ invariant mixed states in $H_n\otimes H_n$ \cite{M. Horodecki1999}:
\begin{eqnarray*}
\rho_{iso}(f)
&=&\frac{1-f}{n^2-1} I_n +  \frac{n^2f-1}{n^2-1} |\psi^+ \rangle
\langle \psi^+|,
\end{eqnarray*}
with $f=\langle \psi^+| \rho_{iso}(f) |\psi^+ \rangle$ satisfying
$0\leq f \leq 1$, $|\psi^+
\rangle=\frac{1}{\sqrt{n}}\sum_{i=0}^{n-1}|ii\rangle$. These states
are shown to be separable if and only if they are PPT, i.e. $f \leq
\frac{1}{n}$. Now we utilize inequality (\ref{ineq nbyn*}) with
$m=n$. It can be verified that the left hand side of inequality
(\ref{ineq nbyn*}) is $(\frac{1-f}{n^2-1})^2$ and the right hand
side of inequality (\ref{ineq nbyn*}) is
$(\frac{n^2f-1}{n^2(n^2-1)})^2$ for $\rho_{iso}(f)$. If we choose
$U=V=I_n$ in inequality (\ref{ineq nbyn}), then the violation of
this inequality for this isotropic state is $-\langle
W^{(n,n)}_{I_n,I_n}\rangle_{\rho_{iso}(f)}=(\frac{n^2f-1}{n^2(n^2-1)})^2-(\frac{1-f}{n^2-1})^2$,
which is  positive if and only if $f>\frac{1}{n}$. Therefore,
inequality (\ref{ineq nbyn*}) can detect all the entanglement of
isotropic states.

Example 3. Werner states are a class of $U \otimes U$ invariant
mixed states in $H_n\otimes H_n$ \cite{werner}:
\begin{eqnarray*}
\rho_{wer}(f) = \frac{n-f}{n^3-n} I_n + \frac{nf-1}{n^3-n}\tilde{V},
\end{eqnarray*}
where $\tilde{V}=\sum_{i,j=0}^{n-1} |ij\rangle \langle ji|$ and
$f=\langle \psi^+| \rho_{wer}(f) |\psi^+ \rangle$, $-1\leq f \leq
1$. These states are separable if and only if they are PPT, i.e. $f
\geq 0$. According to inequality (\ref{ineq nbyn}), we choose
$U=|0\rangle \langle 1| +|0\rangle \langle 1|+\sum_{i=2}^{n-1}|i\rangle\langle i|$ and $V=I_n$,  then
the violation of this inequality is $-\langle
W^{(n,n)}_{U,V}\rangle_{\rho_{wer}(f)}=(\frac{nf-1}{n^3-n})^2-(\frac{f+1}{n(n+1)})^2$,
which is positive if and only if $f<\frac{2-n}{2n-1}$ (See FIG. 1).
Obviously, when $n=3$, Werner state violates this inequality if
$f<-0.2$.
\begin{center}
\begin{figure}[!h]
\resizebox{5cm}{!}{\includegraphics{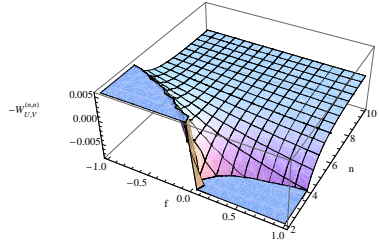}}\caption{The violation value of Werner state
with respect to inequality (\ref{ineq nbyn}) under $U=|0\rangle \langle 1| +|0\rangle \langle 1|+\sum_{i=2}^{n-1}|i\rangle\langle i|$ and $V=I_n$.
}
\end{figure}
\end{center}

In the following, we study the relations between
inequality (\ref{ineq nbyn}) and entanglement distillability.

\begin{theorem}\label{th dis}
A state $\rho\in H_m\otimes H_n$ is distillable if and
only if there exist two projectors $A$ and $B$ mapping  high
dimensional spaces to two dimensional ones such that the restriction
of the state of $N$ copies of $\rho$ to such $2\otimes 2$ subspace
violates inequality (\ref{ineq nbyn}).
\end{theorem}

Proof. Any state $\rho$ is distillable if and only if there exist
two projectors $A$ and $B$ mapping high dimensional spaces to two
dimensional ones such that $A\otimes B\rho^{\otimes N} A\otimes B$
is entangled \cite{M. Horodecki1998}. However, $A\otimes
B\rho^{\otimes N} A\otimes B$ is two-qubit entangled if and only if
there exist $m^N \times m^N $ unitary matrix $U_0$, $n^N \times n^N
$ unitary matrix $V_0$, and nonzero real numbers $a$ and $b$,
$a^2+b^2=1$, such that $U_0\otimes V_0 (a|01\rangle +b|10\rangle)$
is the eigenvector of $A\otimes B(\rho^{\otimes N})^{T_1} A\otimes
B$ with respect to a negative eigenvalue. Taking into account that inequality
(\ref{ineq nbyn}) is the necessary and sufficient condition for
two-qubit entanglement \cite{mj}, we have that $\rho$ is distillable
if and only if one finds two projectors $A$ and $B$ mapping high
dimensional spaces to two dimensional ones such that the restriction
of the state of $N$ copies of $\rho$ to such $2\otimes 2$ subspace
violates inequality (\ref{ineq nbyn}).\qed

Some necessary conditions for distillability have been proposed in Refs. \cite{M.
Horodecki1999,Y. Ou,M. Li}. The distillability
condition we obtained here is both necessary and sufficient and could be
verified experimentally in principle.

Example 4. For non-PPT (NPT) entangled state
\begin{eqnarray*}
\varrho_1&=&\frac{p}{6}(|00\rangle \langle 00| + |01\rangle \langle 01| +|02\rangle \langle 02|+|10\rangle \langle 10|
+|11\rangle \langle 11|+ |12\rangle \langle 12|)-\frac{p}{6}(|00\rangle \langle 12|\\
&&+|01\rangle \langle 12|+|12\rangle \langle 00| +|12\rangle \langle 01| +|10\rangle \langle 11|+|11\rangle \langle 10|)
+ \frac{1-p}{2}(|22\rangle\langle22| +|33\rangle\langle33|),
\end{eqnarray*}
Ref. \cite{mjconcurrence2011} has proved that the distillability of
this state can not be detected by reduction  criterion \cite{M.
Horodecki1999} and the criterion in Ref. \cite{Y. Ou}. The
distillability of this state can not be recognized either by the
result in Ref. \cite{M. Li}. However, setting
$A=|0\rangle\langle0|+|1\rangle\langle1|$ and
$B=(|0\rangle+|1\rangle)(\langle0|+\langle1|)+|2\rangle\langle2|$,
one can verify that $A\otimes B \varrho_1 A\otimes B$ violates
inequality (\ref{ineq nbyn}) with $m=2$, $n=3$ and $U=I_2$,
$V=|1\rangle\langle2|+
\frac{1}{\sqrt{2}}|0\rangle(\langle0|+\langle1|)+
\frac{1}{\sqrt{2}}|2\rangle(\langle0|-\langle1|)$. Therefore our
inequality gives a better recognition of distillability for this case.
\medskip

\section{Comments and Conclusions}

We have provided an inequality in terms of the expectation values of
local observables, each with two possible  measurement outcomes, for
detecting entanglement and distillability for arbitrary dimensional
bipartite systems. Any violation of this inequality implies that the
state being measured is entangled. Moreover, all pure entangled
states violate the inequality, namely the inequality is both necessary and
sufficient in detecting entanglement for bipartite pure states.
As examples we have analyzed the entanglement
detection of the Horodecki's state, isotropic state and Werner
state. It has been shown that the inequality can detect considerable
mixed entangled states. Above all, the inequality is a necessary and sufficient condition of distillability. The results may be used in
experimental entanglement detection and distillability verification.

In addition, since the dimensions of the two subsystems in
inequality (\ref{ineq nbyn}) are arbitrary,  this inequality could
be also used to detect entanglement in multipartite systems: if a
multipartite state is bipartite separable under some partition, then
it fulfills inequality (\ref{ineq nbyn}) under the corresponding
bipartite partition. A multipartite pure state is bipartite
separable if and only if it fulfills inequality (\ref{ineq nbyn}) for all unitary operators $U$ and $V$ under this bipartite
partition, for which the two subsystems may have different
dimensions.

\bigskip
\noindent{\bf Acknowledgments}\,
S. M. Fei acknowledges the support from NSFC under No.
11275131.

\end{document}